\documentstyle[preprint,aps]{revtex}
\tightenlines
\begin{document}
\draft
\title{\bf Ambiguity in the 
evaluation of the effective action
on the cone}
\author{Devis Iellici\footnote{Electronic address: 
devis.iellici@telital.com}}
\address{Dipartimento di Fisica, 
Universit\`a di Trento \\ and 
Istituto Nazionale di Fisica 
Nucleare, Italy.}
\author{Edisom S. Moreira Jr.\footnote{Electronic address: 
moreira@cpd.efei.br}}
\address{Instituto de F\'{\i}sica
Te\'orica, Universidade Estadual Paulista,\\
Rua Pamplona 145, 01405-900, 
S\~ao Paulo, S.P., Brazil.}

\date{July 1998}
\preprint{hep-th/9807184}

\maketitle

\begin{abstract}
An ambiguity in the computation of the one-loop
effective action for
fields living on a cone is illustrated. 
It is shown that the ambiguity arises due to the
non-commutativity of the regularization of
ultraviolet and (conical)
boundary divergencies.
\end{abstract}

\narrowtext 

\section*{INTRODUCTION}

Recently there has been much interest 
in computations of one-loop
effects for quantum fields living in 
backgrounds with conical
singularities, especially in the context of quantum corrections to the
Bekenstein-Hawking entropy 
\cite{thooft,SU}. In this context, conical
singularity methods
\cite{SU,CW94,DO94,CKV94,SOL95,kabat,carteit95,delamy95,LW,FUSOL96,ZCV,ffz96,elirom96,hotta,ffz97}
have been shown to be a powerful tool. 
There is, nevertheless, a controversy
in the literature concerning two 
possible kinds of approaches to computing
one-loop effects on the cone, 
leading to different results (for a
review, see \cite{fro98}).

In the {\em local approach} 
\cite{DO94,CKV94,ZCV}, one starts
regularizing and renormalizing local quantities, 
such as the effective
Lagrangian density or the energy-momentum tensor. 
Global quantities,
for example the effective action, 
are then obtained by integrating the
local ones over the background. 
Local quantities show a
non-integrable singularity at $r=0$, 
where $r$ is the proper distance
from the apex of the cone. 
For a massless field in $D$ dimensions, for
example, simple dimensional considerations 
determine the dependence of the
effective Lagrangian density on $r$ as being $r^{-D}$. 
Therefore integration over the background usually 
requires the introduction of a
cut-off at a proper distance 
$\epsilon$ from the tip of the cone.

In the {\em global approach}
\cite{CW94,SOL95,LW,FUSOL96,ffz96,hotta,ffz97}, integration over
the background is performed before 
ultraviolet regularization,
resulting that the volume cut-off 
$\epsilon$ is not required. At this
point it should be remarked that 
the local and global approaches are
equivalent to each other when 
the background is a smooth manifold.

In the context of quantum fields at 
finite temperature in Rindler
space-time (or near the horizon 
of a black hole), both approaches lead
to a divergent entropy. However 
the origin of the divergences seems
quite different -- in the local approach 
the divergence is associated
with the (divergent) local temperature 
on the horizon \cite{BARB94},
contrasting with the ultraviolet (u.v.) 
nature of the divergence in
the global approach.
Another important difference concerns 
dependence on the temperature
$T$ of the thermodynamical quantities 
\cite{EMP95,ZCV,elirom96}. In
the local approach the free energy, 
at high temperatures, shows a
leading contribution proportional 
to $T^D$, whereas in the global
approach it behaves as $T^2$. 
Finally, in the global approach the
divergences can be renormalized 
into the bare gravitational action
\cite{SU,LW,FUSOL96}, whereas such a
procedure is not possible in the
context of the local approach \cite{phd}.

Considering these facts, it can be 
said that there is an {\em
ambiguity} in the computation 
of one-loop effects on the cone.
As different u.v. regularization 
techniques have been used in the
literature (for example, the 
$\zeta$ function procedure
\cite{ZCV,massive} in the case of 
the local approach, and the
Schwinger proper-time 
\cite{CW94,SOL95,LW,FUSOL96,hotta,ffz97} 
or the
Pauli-Villars \cite{delamy95} 
procedures in the case of the global
approach), it was not clear whether 
the origin of the ambiguity had to
do with their different 
features or some other reason.
In this paper we show that the origin 
of the ambiguity is not in the
u.v. regularization employed 
(since it appears within any
regularization scheme as will be shown), 
but that it arises from a
conflict between regularization 
of u.v. and horizon
divergencies. One may say, 
from the mathematical point of view, that
the conflict is simply due to the 
fact that the u.v. regulator
$\delta$ and the horizon cut-off 
$\epsilon$ appear as in
$\epsilon^2/\delta$, and therefore 
the result depends radically on the
order in which these regulators are sent to zero.

The next section illustrates the ambiguity 
by employing the Schwinger
proper-time regularization. 
A discussion follows where we give a
particle-loop interpretation 
to the two approaches and argue that
the local approach seems to be 
supported by physical considerations.
In Appendixes A and B, 
the ambiguity is considered in the context of a
point-splitting regularization, 
and in a wide class of Schwinger-like
regularizations, respectively.


\section*{LOCAL {\small v.s.} GLOBAL APPROACH}

Let us see how the ambiguity arises. 
For simplicity a scalar field
will be considered, but the argument 
is essentially the same for
higher spins. The starting point 
is a representation of the local heat
kernel for minus the 
Laplace-Beltrami operator $A=-\Delta_\beta$
on the four-dimensional flat cone 
$C_\beta\times R^2$ with metric
$ds^2=r^2d\tau^2+dr^2+dy^2+dz^2$, 
where $\tau\in[0,\beta]$, $r\in R^+$
and $y,z\in R$. The deficit angle of 
the cone is $2\pi-\beta$. In the
context of finite temperature theory 
in the Rindler space-time (or
near the horizon of a black hole), 
$\beta ^{-1}$ is the temperature.
A useful representation 
of the heat kernel is given by
\cite{carslaw19,DO87a,fursaev94,CKV94}
\begin{eqnarray}
K_t^\beta(x,x|A)=
\frac{1}{16\pi^2 t^2}+
\frac{i}{32\pi^2\beta t^2}
\int_\Gamma dw\,
e^{-\frac{r^2}{t}\sin^2\frac{w}{2}}
{\mbox{cot}\,}\frac{\pi w}{\beta},
\label{HKreprE}
\end{eqnarray}
where the contour $\Gamma$ consists 
of two branches intersecting the
real axis very close to the origin. 
One branch goes from
$-\pi+i\infty$ to $-\pi-i\infty$, 
and the other one from $\pi-i\infty$
to $\pi+i\infty$. An advantage of 
this representation is that the
flat-space heat kernel (the first 
term on the right-hand side) appears
isolated from the conical 
singularity contribution. This feature will
show clearly that the ambiguity 
arises due to the presence of a
conical singularity.

The local heat kernel Eq. (\ref{HKreprE}) 
can be integrated over the
background to yield the 
well-known integrated heat kernel
\cite{cheeger83}
\begin{eqnarray}
K_t^\beta(A)=
\frac{{\cal A_H}V(C_\beta)}
{16\pi^2 t^2}
+\frac{{\cal A_H}}{48\pi t}
\left(\frac{2\pi}{\beta}-
\frac{\beta}{2\pi}\right),
\label{HKtraceD4}
\end{eqnarray}
where $V(C_\beta)=\beta R^2/2$ is 
the (infinite) area of a cone of
radius $R$, and ${\cal A_H}$ is 
the (infinite) area of the transverse
dimensions -- the horizon area. 
The above integration has been
performed by using formulas computed 
by Dowker in \cite{DO87a} (see
also \cite{fursaev94}). Defining
\begin{eqnarray}
C_{2n}(\beta):=\frac{i}{4\beta}
\int_\Gamma dw \,
\mbox{cot}
\left(\frac{\pi w}{\beta}\right)
\left(\sin^2\frac{w}{2}\right)^{-n},
\label{ciennealpha}
\end{eqnarray}
one has for $n=1$, $2$
\begin{eqnarray}
C_2 (\beta)=\frac{\pi}{3\beta}
\left(\frac{2\pi}{\beta}-
\frac{\beta}{2\pi}\right)
&\hspace{1.0cm}&
C_4 (\beta)=\frac{1}{15}
C_2(\beta)
\left[\left(\frac{2\pi}
{\beta}\right)
^2+11\right].
\label{C2C4}
\end{eqnarray}

Let us focus on the computation of 
the one-loop effective action for a
neutral massless scalar field on 
the cone (the massive case is treated
in appendix A), noting that the 
ambiguity should also be present in
the computation of other global quantities. 
The effective action is
determined by integrating 
the effective Lagrangian over the
background
\begin{eqnarray*}
W_{\mbox{\scriptsize eff}}=
\int d^4x\sqrt{g}
L_{\mbox{\scriptsize eff}}(x).
\end{eqnarray*}
Since, from dimensional considerations, 
it is expected that
$L_{\mbox{\scriptsize eff}}(x)$ 
behaves as $r^{-4}$, in the
integration over $r$ above one 
introduces a cut-off at a distance
$\epsilon$ from the horizon ($r=0$). 
The effective Lagrangian can be
computed by integrating over 
the proper-time $t$ the product of the
local heat kernel by $t^{-1}$, 
and since it is u.v. divergent
some regularization procedure is required. 
Here we will apply the
Schwinger regularization which is 
widely used in the literature. Note
however that the results hold 
for any regularization procedure (see the
Appendixes).

According to the Schwinger 
regularization procedure, an infinitesimal
u.v. cut-off $\delta$ 
is introduced in the integration over
$t$,
\begin{eqnarray*}
L_{\mbox{\scriptsize eff}}(x)&=&
-\frac{1}{2}\int_\delta^\infty
\frac{dt}{t}K_t^\beta(x,x|A),
\end{eqnarray*}
resulting in
\begin{eqnarray*}
W_{\mbox{\scriptsize eff}}&=&-
\frac{\beta{\cal A_H}}{2}
\int_\epsilon^\infty r\,dr
\int_\delta^\infty
\frac{dt}{t} K_t^\beta(r|A).
\end{eqnarray*}
By using Eq. (\ref{HKreprE}), 
one integrates over $t$ and $r$ (note that
if the regulators $\delta$ and $\epsilon$ 
are not set equal to zero in an
early stage, the order in which 
the integrations are performed is not
relevant), resulting in
\begin{eqnarray}
W_{\mbox{\scriptsize eff}}
=-\frac{{\cal A_H}
V(C_\beta)}{64\pi^2\delta^2}
+\frac{i{\cal A_H}}
{128\pi^2\epsilon^2}
\int_\Gamma dw
\frac{\mbox{cot}\frac{\pi w}{\beta}}
{\sin^{4}\frac{w}{2}}\,
\left[e^{-\frac{\epsilon^2}{\delta}
\sin^2\frac{w}{2}}-1\right].
\label{masterformula}
\end{eqnarray}
Clearly as the regulators appear 
in the ratio ${\epsilon^2}/{\delta}$,
the (regularized) effective action 
depends on the order in which the
regulators are sent to zero. 
Note that the first term on the
right-hand side of 
Eq. (\ref{masterformula}) 
encodes essentially the
flat-space ultraviolet divergence 
(which can be renormalized by
redefining the bare cosmological constant), 
whereas the second term
encodes the conical singularity contribution.
Now, noticing that if the horizon 
cut-off $\epsilon$ is kept fixed then the
conical contribution in 
Eq. (\ref{masterformula}) is u.v. finite,
and observing Eqs. (\ref{ciennealpha}) 
and (\ref{C2C4}), when
$\delta\rightarrow 0$ one is led to
(up to $e^{-\epsilon^2/\delta}$ terms)
\begin{eqnarray}
W_{\mbox{\scriptsize eff}}
&=&-\frac{{\cal A_H}
V(C_\beta)}{64\pi^2\delta^2}
-\frac{{\beta\cal A_H}}{32\pi^2\epsilon^2}
C_4(\beta).
\label{efflocal}
\end{eqnarray}
Apart from the u.v. divergent term 
(which is irrelevant, since it can
be renormalized away) this result 
is exactly that obtained in the 
{\em local approach} by means of 
local $\zeta$ function \cite{ZCV},
dimensional \cite{phd} 
and point-splitting regularization (see
Appendix A). Interpreting 
$W_{\mbox{\scriptsize eff}}/\beta$ as a free
energy, the Planckian term ($1/\beta^{4}$) 
reproduces precisely the
one in the result by Susskind and Uglum 
\cite{SU}, which was obtained
by counting eigenmodes and employing 
a WKB approximation. There is
also agreement with respect to the 
mass corrections ($1/\beta^{2}$)
(see appendix A and \cite{fro98}). 
Nevertheless, Eq. (\ref{efflocal})
leads to a non Planckian 
contribution ($1/\beta^{2}$) in the free
energy, which is absent in \cite{SU}. 
The source of this discrepancy
is not well understood as yet, 
but it seems to be related to the
presence of a conical singularity 
in the procedures yielding
Eq. (\ref{efflocal}), 
and to the very interpretation of
$W_{\mbox{\scriptsize eff}}/\beta$ 
as a free energy in the present
context (see \cite{Zmoretti}, 
and references therein). 
Note that
according to Eq. (\ref{efflocal}), 
u.v. divergences affect only the
temperature-independent contribution 
in the corresponding free energy,
as usually is the case.

The {\em global approach} corresponds 
to removing the horizon cut-off
($\epsilon\rightarrow 0$) from 
Eq. (\ref{masterformula}) keeping the
u.v. regulator $\delta$ fixed, 
resulting in 
\begin{eqnarray}
W_{\mbox{\scriptsize eff}}
&=&-\frac{{\cal A_H}
V(C_\beta)}{64\pi^2\delta^2}
-\frac{{\beta\cal A_H}}
{32\pi^2\delta}C_2(\beta),
\label{effglobal}
\end{eqnarray}
which is in agreement with early results 
\cite{CW94,SOL95,LW,FUSOL96}.

There are some important remarks 
to be made concerning the dependence
of the results on the regularization 
procedure employed. In the local
approach the conical contribution in 
$W_{\mbox{\scriptsize eff}}$ is
independent of the regularization 
employed (see the Appendixes). This is
not the case in the global 
approach, which does depend on the
regularization procedure. 
As is shown in Appendix B, 
in the context of
the global approach, 
the conical contribution 
in $W_{\mbox{\scriptsize
eff}}$ vanishes in both 
$\zeta$ function and dimensional
regularizations, 
whereas it consists of 
an u.v. divergent term
according to the Schwinger 
regularization procedure. 
There is no
contradiction though, once one 
realizes that the divergent term is
proportional to the Seeley-DeWitt 
coefficient 
$A_1=\beta{\cal A_H}C_2(\beta)$ 
[see Eq. (\ref{HKtraceD4})], 
and therefore can be
renormalized away by redefining 
the bare Newton constant in the
gravitational action 
\cite{SU,delamy95,LW,FUSOL96}. 
In the local
approach no such renormalization 
is possible, since the conical
contribution is not proportional 
to geometrical terms which can be
found in the gravitational action. 
This is consistent with the
statement above that, in the local approach, 
all regularizations lead
to the same result. 
These facts illustrate 
the different physical
nature of the local and global results. 
In the context of the 
computation of quantum 
corrections to the Bekenstein-Hawking 
entropy, it follows that in the 
global approach the entropy remains
in the form $S={\cal A_H}/4G$, 
which is finite even after quantum 
corrections have been considered, 
provided that $G$ 
is the renormalized Newton constant
\cite{SU}.  In the local approach, 
on the contrary, quantum 
corrections result in a divergent 
total entropy \cite{thooft,BARB94}.

\section*{DISCUSSION}

It is clear that the ambiguity considered 
in this paper arises due to
the  fact  that  the  order  in which u.v. and horizon divergences are
isolated (regularized) matters. Let us discuss on further the subject.

According to a path integral interpretation associated with the heat
kernel, the one-loop effective action is obtained by summing over all
particle loops. In the local approach, one considers a particular point
and the one-loop contribution to the effective Lagrangian is obtained
by summing over all loops starting and ending at that point. The
effective action is then obtained by integrating the effective
Lagrangian over the space. In the global approach, instead, one sums
over all possible closed loops without having to consider local
quantities.

In this scenario, u.v. divergences originate from infinitesimal
loops about to contract to a point. There are two kinds of loops on a
cone. Loops that do not wind around the conical singularity probe only
flat space. Infinitesimal loops of this kind give essentially the
usual u.v. divergence contributions in flat space, which is the same
in both approaches. They correspond to the first term on the
right-hand side of Eqs. (\ref{efflocal}) and (\ref{effglobal}). Loops
of the other kind are those that wind around the conical singularity.
They detect the conical singularity and give rise to
the second terms on the right-hand side of Eqs. (\ref{efflocal}) and
(\ref{effglobal}). In the local approach, these loops cannot be
contracted to a point and so they do not give u.v. divergences
in the local quantities (they are responsible for the conical part in
Eq. (\ref{efflocal}) which is, in fact, u.v. finite). However, by
considering points closer and closer to the apex of the cone, one
picks up contributions corresponding to
smaller and smaller loops encircling
the singularity, and which give rise to the divergence as
$\epsilon\rightarrow 0$. In the global approach, on the contrary,
by summing over all loops one considers also infinitesimally small
loops winding around the singularity. Thus an u.v. divergent
contribution, which depends on the deficit angle, arises
in Eq. (\ref{effglobal}). It is then clear that horizon divergences
in the local approach and u.v. divergences in the global approach
are different manifestations of the same phenomenon.

At this point an inescapable question is posed:
which one of these conical contributions is the correct one?
Mathematically, the origin of the ambiguity is clear, resulting that
both approaches seem to be equally correct. Therefore the answer to
the question above should come from physical arguments. From this
point of view the result given by the local approach seems to be
supported by the fact that, before determining global quantities, one
would like to know corresponding local quantities (such as the
energy-momentum tensor, for example). In order to be meaningful, these
local quantities should have their u.v. divergences appropriately
regularized. It is important to stress that no such calculation is
possible in the context of the global approach, 
since u.v. regularization is performed
only after integration over the background.

There are also some facts supporting the local approach.
Statistical mechanics models for quantum fields vibrating
near the horizon, e.g., the brick wall model \cite{thooft,SU},
fix a Planckian behavior for the free energy ($1/\beta^{4}$).
As mentioned previously, the corresponding local approach result
fits this behavior, at least at high temperatures.
According to the global approach the free energy has a non Planckian
behavior ($1/\beta^{2}$). Furthermore 
one expects the divergence in the free energy 
to be related with the divergence in the local temperature on the 
horizon \cite{BARB94}. This correspondence holds 
in the local approach result.

Before closing, we would like to comment on some previous works
\cite{fro98,rom96,ant97}. The authors of \cite{fro98} mention that the local
approach (roughly speaking local and global approaches correspond,
respectively, to ``volume cut-off'' and ``ultraviolet limit'' in the
terminology of \cite{fro98})
does not correspond
to a complete theory on the cone for the following fact. Since the
integration over $r$ stops at an infinitesimal distance $\epsilon$
from the horizon ($r=0$), and the quantities in the local approach
carry the cut-off $\epsilon$, in the local approach one is in fact
working in an ``incomplete background'' -- the horizon (apex of the
cone) is missing. That is not quite correct. In this paper we have
shown that in both approaches one uses the same heat kernel, which is
well defined on the whole cone and built up with eigenmodes which are
required to be finite on the horizon. By stopping the integration over
$r$ at a certain distance from the horizon one is simply isolating the
horizon divergences, and not truncating the cone as the authors of
\cite{fro98} seem to suggest.

Entropy corresponding to the effective actions considered in this paper
is often identified (for $\beta=2\pi$)
as quantum correction to  the Bekenstein--Hawking entropy of a Schwarzschild
black hole. However we should recall that such effective actions
have been obtained in Rindler spacetime. When compared with
results obtained by working in black hole background itself (see,
for example,
\cite{rom96} where WKB approximation is applied in the context of 
the brick wall model \cite{thooft}, leading to an expression for the
effective action involving the Epstein--Hurwitz $\zeta$ function),
they present the correct leading
divergence on the  black hole horizon
[see, e.g., Eq. (\ref{eaction2})], but fail to reproduce a {\it massless}
logarithmic (subleading) divergence which arises due to the non trivial
topology of the black hole \cite{sol95b}. In semiclassical approaches
topology manifests itself when integration
over the background is performed \cite{mor99}.

It would be worth investigating the issues considered here
in the context of the approach in \cite{ant97}, where
a method for determining heat kernel involving a partial resummation of the 
Schwinger--DeWitt series is presented.

\section*{ACKNOWLEDGMENTS}

D.I. is grateful to Emilio Elizalde, 
Valter
Moretti and, in particular, 
Sergio Zerbini for useful discussions
and suggestions. E.M.
is grateful to George Matsas for 
valuable discussions and to
FAPESP (through Grant No. 96/12259-1) 
for financial support.

\section*{APPENDIX A:
Point-Splitting Procedure}

In this appendix we show how the ambiguity arises when a
point-splitting regularization procedure is used. 

In the following we sketch the evaluation of
\begin{eqnarray}
L_{\mbox{\scriptsize eff}}(x)&=&
-\frac{1}{2}
\lim_{x' \to x}
\int_0^\infty \frac{dt}{t}K_t^\beta(x,x'|A)
\label{rregelagrangian}
\end{eqnarray}
in an $N$ dimensional cone and 
then the associated effective action.
The heat kernel for a massive scalar 
field can be read from the integrand
of the proper time representation 
of the Feynman propagator in 
\cite{moreira95},
\begin{eqnarray}
K_t^\beta(\Delta)=
\frac{2\pi}{\beta (4\pi t)^{N/2}}
e^{-r^{2}/2t-m^{2}t}\sum_{n=-\infty}^
{\infty}I_{2\pi |n|/\beta}
\left(r^{2}/2t \right)
e^{i2\pi n\Delta/\beta}. 
\label{pro3}
\end{eqnarray}
Here $\Delta:=\tau-\tau'$ and 
the coordinates of points $x$ and $x'$ 
have correspondingly been identified,
except the angular coordinates. 
Using \cite{pru86} one finds
\begin{eqnarray}   
&&\int_0^\infty \frac{dt}{t}K_t^\beta(\Delta)=
\frac{2\pi}{\beta (2\pi^{1/2})^{N}r^{N}}\times
\nonumber
\\
&&\quad\qquad\sum_{n=-\infty}^{\infty}
e^{i2\pi n\Delta/\beta}
\left\{\frac{\Gamma\left[N/2+\nu_n\right]
\Gamma\left[(1-N)/2\right]}
{\pi^{1/2}\Gamma\left[(2-N)/2+\nu_n\right]}
\nonumber\right.\times
\\
&&\quad\qquad\left. {}_{1}F _{2}\left[\frac{1-N}{2};\frac{2-N}{2}-
\nu_n ,
\frac{2-N}{2}+\nu_n ;(mr)^{2}\right]\nonumber\right.
\\
&&\quad\qquad\left.+\;2^{-2\nu_n} 
(mr)^{N+2\nu_n}
\frac{\Gamma\left[-N/2-\nu_n\right]}
{\Gamma\left[1+\nu_n\right]}\nonumber\right.\times
\\
&&\left.\quad\qquad{}_{1}F_{2}\left[\frac{1}{2}+\nu_n ;
\nu_n+\frac{N+2}{2} ,
1+ 2\nu_n;(mr)^{2}\right]
\right\},
\label{del}
\end{eqnarray}
where $\nu_n :=2\pi|n|/\beta$ and
${}_{1}F_{2}\left[a ;b,c;z\right]$ 
denotes the generalized
hypergeometric function \cite{sla66} 
which converges for all values of
$z$. Keeping mass corrections in 
Eq. (\ref{del}) up to second order in
$mr$, Eq. (\ref{rregelagrangian}) yields
\begin{eqnarray}
L_{\mbox{\scriptsize eff}}(r)=
\frac{i}{(1-N)r^{2}}
\lim_{\Delta\rightarrow 0}
\left[\partial_{\Delta}^{2}+
\left(\frac{N-2}{2}\right)^{2}+
\frac{1-N}{2}(mr)^{2}\right]
D_{{\cal F}}(\Delta),
\label{elagrangian}
\end{eqnarray}
where $D_{{\cal F}}(\Delta)$ 
is the massless Feynman propagator
\cite{moreira95,bre98}. When $N=4$, 
expansion of $D_{{\cal F}}(\Delta)$
in powers of $\Delta$ is given by 
\begin{eqnarray}
D_{{\cal F}}(\Delta)=
\frac{i}{8\pi^{2}r^{2}}
\left[-\frac{2}{\Delta^{2}}
-\frac{2\pi^{2}}{3\beta^{2}}-
\frac{2\pi^{4}}{15\beta^{4}}\Delta^{2}
+{\cal O}(\Delta^{3})\right].
\label{Feynman}
\end{eqnarray}
Inserting Eq. (\ref{Feynman}) in Eq. (\ref{elagrangian}),
the ultraviolet divergences in the limit
$\Delta\rightarrow 0$ 
may be eliminated by subtracting
the contribution corresponding to
some particular $\beta=\beta_{o}$,
\begin{equation}
L_{\mbox{\scriptsize eff}}(r)=
\frac{1}{24\pi^{2}r^{4}}\left[f(\beta)-
f(\beta_{o})\right],
\label{relagrangian}
\end{equation}
with
$f(\beta):=-4\pi^{4}/15
\beta^{4}-2\pi^{2}/3\beta^{2}
+\pi^{2}(mr)^{2}/\beta^{2}$.
The effective action is obtained 
by integrating the effective
Lagrangian over the background
\begin{equation}
W_{\mbox{\scriptsize eff}}=
{\cal{A_H}}\int_{0}^{\beta}d\tau
\int_{\epsilon}^{R}  dr\ r\   
L_{\mbox{\scriptsize eff}}(r),
\label{eaction1}
\end{equation}
where $\epsilon$ and $R$ are 
the horizon and volume cut-offs
respectively. Finally replacing 
Eq. (\ref{relagrangian}) in
Eq. (\ref{eaction1}) one finds
\begin{eqnarray}
W_{\mbox{\scriptsize eff}}=
-\frac{{\cal A_H}\pi^{2}}{180\epsilon^{2}
\beta^{3}}-
\frac{{\cal A_H}}{72\epsilon^{2}
\beta}
-\frac{{\cal A_H}m^{2}}{24\beta}
\ln\ m\epsilon+C_{\beta_{o}}\ \beta,
\label{eaction2}
\end{eqnarray}
where $C_{\beta_{o}}$ denotes 
a constant  which ensures that 
$W_{\mbox{\scriptsize eff}}=0$ 
when $\beta=\beta_{o}$.
In Eq. (\ref{eaction2}) the upper 
cut-off contribution has been omitted.

When $m=0$ Eq. (\ref{eaction2}) 
is in agreement with Eq. (\ref{efflocal}),
and the mass correction is 
in agreement with that computed by other
methods \cite{byt96,massive}. 
It is worth remarking that the result
above differs from that obtained 
by Dowker \cite{DO94},
where the expectation value of the 
energy density is integrated
over the background to give 
the internal energy. 
The difference 
is in the non Planckian 
massless term in the corresponding
free energies. As mentioned previously, 
there has not appeared yet a
satisfactory explanation for these 
low temperature discrepancies
(see \cite{Zmoretti,phd}).

If in Eq. (\ref{eaction1}),
by setting $\epsilon=0$ and $R=\infty$,
the integration over the background 
is performed before the 
integration over $t$, 
then one is 
led to the global approach result 
\cite{kabat}.
Note that in this case the splitting
$\Delta=\tau-\tau'$ is not enough to 
handle u.v.
divergences, and a further regulator 
(e.g., the proper time
cut-off $\delta$) is required.

\section*{APPENDIX B: 
Generalized Schwinger Regularization}
This appendix considers the ambiguity
in the context of a wide class 
of Schwinger type regularizations.
In general the effective action
can be computed from the heat-kernel as follows
(see, e.g., \cite{dokcrit76,ball,report})
\begin{eqnarray*}
W_{\mbox{\scriptsize eff}}&=&-\frac{1}{2}
\lim_{\delta\rightarrow 0}
\int_0^\infty\frac{dt}{t}\rho(\delta,t)K_t(A),
\end{eqnarray*}
where the regularizing function $\rho(\delta,t)$ 
has to satisfy some
requirements. For example, 
it is assumed that 
$\lim_{\delta\rightarrow 0}
\rho(\delta,t)=1$ and that 
(for sufficiently large $\delta$)
$\rho(\delta,t)$ has to 
regularize the divergence at $t=0$.
Among possible
regularizing functions 
one has the proper-time cut-off, 
$\rho(\delta,t)
=\Theta(t-\delta)$ 
(considered previously in this work), 
$\zeta$ function, 
$\rho(\delta,t)=\frac{d}{d\delta}
[t^\delta/\Gamma(\delta)]$, 
dimensional regularization, 
$\rho(\delta,t)=(4\pi t)^{-\delta}$,
and Pauli-Villars regularization, 
$\rho(\delta,t)=(1-e^{- t/\delta})^3$
(in four dimensions). Effective
actions computed using different regularizing functions 
differ by terms which are divergent as $\delta\rightarrow 0$,
but proportional to the Seeley-DeWitt coefficients $A_0$, $A_1$ 
and $A_2$. Therefore these terms can be renormalized away
by redefining the bare coupling constants in the 
gravitational action \cite{birrel}.

Let us then consider a cone as background. 
Employing an arbitrary regularizing function
(flat space contributions, which can be renormalized
away, will be dropped) Eq. (\ref{masterformula}) becomes
\begin{eqnarray}
W_{\mbox{\scriptsize eff}}&=&
-\frac{i{\cal A_H}}{128\pi^2}
\int_\Gamma dw 
\frac{\mbox{cot}\frac{\pi w}{\beta}}
{\sin^{2}\frac{w}{2}}\int_0^\infty 
\frac{dt}{t^2}\rho(\delta,t)
e^{-\frac{\epsilon^2}{t}
\sin^2\frac{w}{2}}.
\label{grandmasterformula}
\end{eqnarray}
In the local approach 
one sets $\delta\rightarrow
0$, so that $\rho(\delta,t)\rightarrow 1$.
After integrating over $t$ one is left with Eq.
(\ref{efflocal}), whichever regularizing
function one chooses. In the global approach, on the
other hand, one sets $\epsilon
\rightarrow 0$ first, so that 
\begin{eqnarray}
W_{\mbox{\scriptsize eff}}&=&
-\frac{\beta{\cal A_H}}{32\pi^2} C_2(\beta)
\int_0^\infty \frac{dt}{t^2}\rho(\delta,t).
\label{effglobal2}
\end{eqnarray}
The integral over $t$ depends on the regularizing function. It is
divergent in some regularizations (proper-time cut-off, Pauli-Villars
regularizations) and vanishing in others (dimensional,
$\zeta$ function regularizations). But as the result is proportional
to the Seeley-DeWitt coefficient $A_1=\beta{\cal A_H} C_2(\beta)$, it
can be renormalized away into the bare gravitational action. This fact
ensures equivalence of different regularizing functions.

\end{document}